# Distributed resource allocation in cognitive radio networks with a game learning approach to improve aggregate system capacity

*José Ramón Gállego, María Canales, Jorge Ortín*

## AUTHORS AFFILIATION

Aragón Institute of Engineering Research, I3A, University of Zaragoza.

C\ María de Luna, 3, 50.018, Zaragoza (Spain)

## CONTACT INFORMATION

jrgalleg@unizar.es (Corresponding author)

mcanales@unizar.es

jortin@unizar.es

# ABSTRACT

*This paper presents a game theoretic solution for joint channel allocation and power control in cognitive radio networks analyzed under the physical interference model. The objective is to find a distributed solution that maximizes the network utility, defined with different criteria, with limited information. The problem is addressed through a non-cooperative game based on local information. Although the existence of a pure Nash equilibrium cannot be assured for this game, simulation results show that it exists with high probability and with a performance similar to that of a potential game, where each player requires overall network information. The obtained results are compared with a centralized heuristic genetic algorithm to show the correctness of the proposals. From this point, utility functions for the local game are modified to restrict the transmitted power to drive the solution to a more cooperative approach. To overcome the convergence limitations of the local game, no-regret learning algorithms are used to perform the joint channel and power allocation. These algorithms provide stable mixed strategies in any scenario with even better global performance. This opens an interesting perspective to develop realistic protocols based on the modeled interactions and increases the adaptability to perform efficient opportunistic spectrum access.*

# 1. INTRODUCTION

The expansion of wireless applications and mobile devices during the last years has enabled the development of flexible network architectures and efficient technologies to exploit the increasingly scarce radio resources. In this context, autonomous, self-configuring multihop networks present a versatile solution to provide broadband services with infrastructure-less deployments, based on a decentralized management. Furthermore, their intrinsic adaptability and resilience can be enhanced with Cognitive Radio technology (CR) [1]. CR devices are able to reconfigure their transmission parameters (frequency band, transmission power, waveforms) depending on the environment conditions. This emerging technology enables to perform opportunistic spectrum access to unlicensed bands or to spectral holes in the licensed ones. Namely, in a cognitive radio network, users who do not own spectrum license are known as secondary or unlicensed users, and the spectrum license holders are known as primary or licensed users. According to the spectrum bands that secondary users are accessing, spectrum sharing and allocation schemes can be divided into two types [4]: Spectrum sharing among the secondary users who access unlicensed spectrum bands (open spectrum sharing) and spectrum sharing among the secondary users and primary users in licensed spectrum bands (licensed spectrum sharing). In this latter case secondary users are allowed to access the licensed bands as long as they will not generate harmful interference to the primary users. In both cases, devices in a cognitive network should adaptively adjust their operating parameters based on the interactions with the environment and other devices in the network to improve the system capacity.

Specifically, within the context of self-configuring multihop networks [2], one of the main research challenges to perform an efficient and distributed radio resource management is the joint selection of frequency channel and transmission power. In this case, overall information about the environment is not assured at each user, so they may have to take decisions based on limited environmental information. In addition, individual requirements can be in contrast to optimizing the overall network performance. Therefore, the appropriate rules to drive the reconfiguration and adaptation process have to solve this trade-off.

To tackle these challenges, game theory has recently received an increasing interest in the context of cognitive networks [3], [4]. Game theory is a mathematical tool that analyzes the strategic interactions among multiple decision makers. This characterization of the problem facilitates to evaluate an expected performance, which can provide interesting results to understand the operation of these networks and to guide the required protocols design. Moreover, in the context of self-configuring wireless networks, where centralized control is not available, non-cooperative games can provide efficient and distributed solutions for radio resource management [4]. In general, this kind of analysis focuses on modeling the players' interaction with specific types of games and, according to the game settings, deriving the equilibrium criteria (e.g. Nash Equilibrium – NE – strategies profile where no player has anything to gain by unilaterally deviating). Games that converge to stable equilibrium solutions, such as potential games [5], can provide predictive performance when users rationally play according to the specified rules. On the other hand, these guarantees can come at the expense of using a complete knowledge of the environment, which makes the distributed operation more complex. Therefore, defining appropriate decision rules dependent only on local information demands increased efforts.

Parallel to game theory, research in Multi-Agent Learning (MAL) studies the interaction of agents reactive to its changing environment, which, from the perspective of a single agent, includes the behavior of the opponents. By means of an algorithm that learns from previous experience (measured as a reward), agents adjust their behavior. Although the MAL perspective comes from a different approach in the artificial intelligence community, it can be related to game theory [6]. In fact, there are many studies about computing equilibria form a learning perspective and analyzing the success of the defined rules (convergence, regret) [7]. In this context, regret minimization approaches [8] have some unique properties that distinguish it from other MAL works (model-based as fictitious play or model-free as reinforcement learning). Instead of trying to reach an equilibrium, these techniques seek to minimize regret: doing as well as playing a best response to the historical frequency distribution (universal consistency [9]). Against unknown opponents utilizing strategies difficult to model, universally consistent policies may also be a good choice for guiding the play. As an example, myopic players in a repeated game that observe the actions, but not the payoffs of the other players can learn and adapt the strategies in each stage, converging to alternative solutions even beyond the NE (coarse correlated equilibria [10], [11]).

The objective of this paper is to find an efficient distributed joint channel and power allocation in a multihop cognitive network. To this purpose, we model the system as a non-cooperative game and propose different utility functions with the ultimate goal of maximizing some overall network utility without a complete knowledge of the environment. The definition of this network utility can be highly dependent on the applications' requirements. For this reason, we have considered two possibilities: the aggregate capacity in bps of all the established links (elastic applications) and the total number of established links (fixed rate inelastic applications). The proposals are analyzed and evaluated with the physical interference model [12]. Under this model, a link can only be established if the received Signal to Interference-Noise Ratio (*SINR*) is higher than a predefined threshold.

In this context, we propose a repeated game characterized by the limited information used to drive the decision process. The individual goals of players are initially defined in terms of its own utility (capacity in bps or link activation) subject to the *SINR* restriction. The use of local information restricts the properties of the game, and the existence of a NE cannot be assured. However, as the presented results show, following a best response criterion, a NE is reached most of the cases and the overall performance of these NEs is comparable to a reference potential game that considers global information and exhibits convergence to the set of Nash equilibria. Upon the basis of these results, the designed utility functions of the local game are modified to restrict the transmitted power, driving the solution to a more cooperative approach that improves the global performance. In addition, to overcome the convergence limitations of the local game, increased with the inclusion of restrictions in the transmitted power, no-regret learning algorithms based on [13] are included in the work. These algorithms can obtain stable mixed strategies with a suitable global performance.

The joint channel and power allocation problem for cognitive radio networks have been already studied under a game theoretical and learning perspective [14]-[18]. However, these works are based on a cooperative approach through a potential game where each player requires global information about the remaining players in the network. In some cases [14], [17], games with local information are also proposed, but the obtained performance is significantly worse than that achieved with the potential game. On the other hand, the game with local information proposed in this paper, with myopic decisions subject to *SINR* restrictions, provides similar performance to the equivalent potential game under different utility definitions dependent on applications' requirements. In addition, the efficiency of the obtained equilibria of both cooperative and local approaches is often not taken into account. In this context, it is known to be hard to find the channel and power allocation to obtain the maximum throughput in wireless networks. Since the problem is usually NP-hard [19], its analysis is typically tackled with heuristic algorithms. Thus, in this paper, to provide a reference that confirms the validity of the proposals and the quality of the obtained equilibria, we compare the obtained results with a centralized meta-heuristic genetic algorithm [20].

The remaining of the paper is organized as follows: Related work and main contributions are presented in Section 2. In Section 3 the system model is described. In Section 4, the proposed game theoretic and learning approaches are explained. Section 5 shows the simulation framework and the obtained results. Finally, some conclusions and prospects are provided in Section 6.

## 2. MOTIVATION

### 2.1 Related Work

Game theory has received a lot of attention to analyze wireless networks [21] and more specifically, cognitive radio networks [4] in the last years, analyzing multiple aspects, such as routing [22] or access point selection in wireless mesh network [23]. In this section we provide an overview of those contributions more related to our work, in particular focused on channel allocation and power control for multihop networks.

Channel allocation for multi channel multi radio wireless networks is studied from a game theoretical perspective in [24]-[26]. These works analyze the channel assignment under a simplified propagation model with fixed transmission and sensing ranges. In [24] and [25] all the nodes are assumed to be in a single collision domain, i.e., all the links are in the sensing range of each other, whereas in [26] the analysis is extended to a multiple collision domain, where each link need not be in the sensing range of every other link in the network. Transmission power, and therefore, transmission and interference ranges are fixed and aspects such as the quality or capacity of the links, measured through the received *SINR* are not taken into account in these works.

Closer to our work is the game theoretic framework to perform channel allocation in cognitive radio links proposed in [14]. In this case, the interference generated by each link is taken into account in the definition of the utility function of each player. Both cooperative and selfish strategies are analyzed by means of a potential game and a no-regret algorithm respectively. However, power control is not considered and all the links are assumed to transmit with fixed and equal power. In addition, the proposed selfish strategies provide significant worse performance than the potential game. The same authors include power control in [15], but only for the cooperative strategy based on a potential game. Based on these works, a similar utility function for the cooperative strategy with the inclusion of power control is proposed in [16]. Nevertheless, in all these cases, the definition of the utility function does not directly represent the application requirements according to the received *SINR* (capacity in bps, link activation if the received *SINR* is higher than the *SINR* threshold) and no reference about the goodness of the obtained results is provided. A similar approach is analyzed in [17]. There, instead of allocating a channel for each link, several channels with different transmission powers can be jointly allocated for the same link. The utility functions are defined taking into account the capacity in bps of the links obtained for a given *SINR* according to the Shannon-Hartley theorem (as described in section 3). As in [14], both cooperative and selfish strategies are studied. In this case, the cooperative game clearly outperforms the selfish one, since in the latter each player will try to occupy all the available channels at the maximum transmission power. However, as in [14], [15] and [16] the cooperative game, which provides better performance, assumes that each player has global information of the strategies of all the remaining players in the network. Obtaining this information in a real deployment implies very high costs of signaling overhead [27], making the solutions not scalable. Again, no results about the optimality of the obtained performance are provided. In [18], a similar potential game is proposed to maximize the throughput of a cognitive network performing a joint channel and power allocation for secondary users. Players again require global information about the remaining players in the network. In that work, the analysis of more practical scenarios reducing the amount of information to be exchanged is left as a future line of work.

In [28] we proposed a potential game for joint channel and power allocation which did not require global information and we provided a reference for its performance by comparing it with a meta-heuristic genetic algorithm [20]. However, this analysis was based on the simplified protocol interference model [12]. Under this assumption, the required information to define a potential game was limited to a reduced number of interfering links (those in the interference range), but the proposed potential game cannot be directly applied to the more realistic physical interference model.

Under this latter interference model, the definition of a potential game requires each link has global information about the remaining links in the network, which makes the solution not be scalable. As a consequence, other kind of games, based on local information, should be considered. However, the limited information about other players' strategies can limit the convergence properties of the game. In these cases, even though a pure NE cannot be found, other solutions, even with better global performance, can be obtained using different approaches. As stated before, within the broad context of multi-agent learning, regret minimization techniques [8] offer an interesting solution in scenarios where reaching a Nash equilibrium is not possible. In the considered channel and power allocation issue, each player repeatedly faces a decision problem with a set of possible actions, each one with a different reward, which varies in the successive iterations of the decision problem. By using no-regret learning algorithms, players acquire knowledge about their environment as the game is played iteratively, with the aim of maximizing average rewards. No-regret learning algorithms use a regret measure, which is obtained comparing the performance of the learning algorithm with a set of alternative strategies. The definition of this set characterizes the different types of no-regret algorithms, such as no-external-regret and no-internal-regret. In [29], a general class of no-regret learning algorithms called ϕ-no-regret learning algorithm is presented and related to game theoretic equilibria. This generalization covers from no-external-regret learning to no-internal-regret learning.

A learning algorithm has no-external-regret (or universal consistency [9]) if it provides better reward than the utilities obtained by playing any other fixed sequence of decisions. This guarantees that the empirical distribution of play converges to the coarse correlated equilibrium set [8], [11]. A probability distribution over the set of joint actions of all the players is a coarse correlated equilibrium if each player's expected reward is at least as high as its expected reward if it were to deviate and play an arbitrary action, while the other players follow the outcome specified by the given joint probability distribution.

On the other hand, no-internal-regret algorithms provide stronger guarantees, since instead of comparing against a set of fixed actions, they compare the reward provided by the algorithm with the rewards obtained by a modified algorithm, which consistently replaces one action by another, for all possible actions. In this case, the empirical distribution of game converges to the correlated equilibrium set [8], [30]. A probability distribution over the set of joint actions of all the players is a correlated equilibrium if none of the players obtain higher reward by playing a different action from the one given by the probability distribution after his part has been revealed, which is a more restrictive condition. In fact, every correlated equilibrium is a coarse correlated equilibrium. Both coarse correlated equilibrium and correlated equilibrium are generalizations of the Nash equilibrium concept which take into account the correlations among the players. The difference between both coarse correlated and correlated equilibrium is the ability of a player in a correlated equilibrium to simultaneously consider multiple alternatives, conditioned on its action profile.

A no-external-regret algorithm (Hedge algorithm, Freund and Schapire - [31]) have been already used in the context of channel allocation in previously described works [26] and [14], to analyze scenarios where limited information is available for the players and convergence to a NE cannot be guaranteed. In a slightly different scenario, but within the context of cognitive radio, in [32], the no-internal-regret algorithm proposed in [30] enables secondary users to adjust their transmission probabilities over the available channels not used by primary users, so that collisions are avoided.

## 2.2 Main Contributions

This paper focuses on channel allocation and power control strategies in distributed wireless networks using game theory. Unlike previous works in this topic [14]-[18], [28], the analysis is performed under a more realistic physical interference model which considers that a link can only be established if the received *SINR* is higher than a predefined threshold. In addition, the problem is studied with different utility definitions dependent on applications' requirements.

Potential games are often used to model similar problems since they ensure convergence to pure NE, which in addition can be global maximizers of the potential function defined for the game. However, the main problem of these games is that players usually require overall information about the remaining players in the network, making the solution not scalable [14]-[18]. The potential game presented in [28] does not require global information, but it cannot be applied to the physical interference model considered in this work, since it was developed for the less realistic protocol interference model. Although some local games have been proposed [14], [17], the obtained performance is clearly worse than that achieved with the corresponding potential games. One of the main contributions of this work is the proposal of a game with local information that provides similar performance to that of a potential game. The main distinguishing features of this local game are the inclusion of *SINR* and power constraints to reduce the selfishness of the players without requiring explicit coordination. Moreover, its performance is not only compared to a potential game designed for this scenario, but also with a heuristic genetic algorithm. A preliminary version of this local game without power constraints was presented in [33].

In addition, in this paper we demonstrate that existence of a pure NE cannot be assured for this local game. Consequently, to solve these convergence limitations we include no-regret learning algorithms capable of obtaining stable mixed strategies. We have focused on no-external-regret algorithms due to their lower computational cost, which can be a critical issue in the context of cognitive radio. We consider this approach as a first step leaving the study of no-internal-regret algorithms as an ongoing line of work. Specifically, we apply Hedge algorithm and a no-external adaptation proposed by Hart and Mas-Colell [30] for our considered joint channel and power allocation problem and analyze their performance in this scenario.

## 3. SYSTEM MODEL

There are two widely used models to characterize interference relationships in a wireless network, the protocol model, also known as the unified disk graph model and the physical model, also known as the *SINR* model. Under the protocol model, a successful transmission occurs when a node is within the transmission range of its transmitter and it is outside the interference ranges of other transmitters. The setting of the transmission range is based on a Signal-to-Noise-Ratio (*SNR*) threshold, whereas the setting of interference range is rather heuristic. Although this model does not accurately reflect the interference problem in wireless networks, it is usually applied to develop protocols and algorithms due to its lower computational complexity. In addition, in [34] it is shown that with a suitable election of the interference range, it can provide valid preliminary results, as we show in [28]. On the other hand, under the physical model, the interference produced by all the transmitters is taken into account and treated as noise. This approach highly increases the computational complexity of the analysis, but it also provides more realistic and reliable results. For this last reason, in this paper we assume this interference model, as we describe next.

Given a directional link $i$ between a pair of nodes $(i_{TX} \rightarrow i_{RX})$, the channel gain from transmitter ($i_{TX}$) to receiver ($i_{RX}$) is defined as $g_{i,i} = d_{i,i}^{-\gamma}$, being $d_{i,i}$ the distance from $i_{TX}$ to $i_{RX}$ and $\gamma$ the path loss index. Similarly, $g_{i,j} = d_{i,j}^{-\gamma}$ represents the channel gain from the transmitter of link $i$ ($i_{TX}$) to the receiver of link $j$ ($j_{RX}$).

The maximum transmission power is set to $P_{max}$. Since typically the actual transmission power can only be set to a finite number of levels, it is discretized into $Q$ levels, equispaced between 0 and $P_{max}$.

There are $F$ non-interfering channels or frequency bands in the network. All the users in the system model are unlicensed secondary users trying to access these generic frequency bands and all of them have the same rights in using the available ones. To model in a simplified way the presence of other kind of interference sources or users (primary licensed users or even non cognitive unlicensed secondary users), it is assumed that some of these $F$ bands are not sensed as available for the modeled users. The set of available bands depends on the geographic location of the users, to introduce space variability in the resource availability.

We assume each link $i$ only uses a channel $f_i$. Under this model, a transmission is successful if the *SINR* at the receiving node is higher than a certain threshold $\alpha$, i.e, if (1) is fulfilled:

$$SINR_i = \frac{p_i \cdot g_{i,i}}{P_N + \sum_{j \in L_T, j \neq i, f_i = f_j} p_j \cdot g_{j,i}} \geq \alpha \qquad (1)$$

where $p_i$ is the power assigned to link $i$, $f_i$ is the channel used for link $i$, $L_T$ is the set of links in the network and $P_N$ is the background noise power.

The design goal is to establish a set $L_T$ of $N$ links between pairs of nodes with the highest network utility. In order to define this network utility, the Shannon-Hartley theorem provides an upper bound of the maximum available capacity for a link $i$ in an AWGN (Additive White Gaussian Noise) channel, $C_i = w_{f_i} \cdot \log_2(1 + SINR_i)$, where $w_{f_i}$ is the bandwidth of channel $f_i$. Under these assumptions, and taking into account the constraint given by (1), the total system capacity can be estimated as:

$$\lambda_i = \begin{cases} 1 & \text{if } SINR_i \geq \alpha \\ 0 & \text{if } SINR_i < \alpha \end{cases} \qquad (2)$$

$$NU = \sum_{i \in L_T} \lambda_i \cdot C_i = \sum_{i \in L_T, SINR_i \geq \alpha} w_{f_i} \cdot \log_2(1 + SINR_i) \qquad (3)$$

However, in a real system, the transmission rate is given by a predefined set of discrete values, basically depending on the modulation and channel coding schemes that guarantee a bit error rate for the actual *SINR*. A simplified way of introducing this effect in the proposed model is applying the Hartley's law, which sets the maximum capacity for a given number of levels in the modulation, $M_i$:

$$C_i \leq 2 \cdot w_{f_i} \cdot \log_2(M_i) \qquad (4)$$

Considering the upper bound given by (4) and the Shannon capacity, and assuming the number of modulation levels is a power of two, the maximum achievable $M_i$ for a given $SINR_i$ can be obtained as:

$$M_i = \lfloor \sqrt{1+SINR_i} \rfloor \qquad (5)$$

The minimum value for $M_i$ is $M_{min}$ = 2, corresponding to a binary modulation. Arbitrarily fixing a $M_{max}$, a fixed set of capacities for each link is defined. This leads to the following network utility:

$$NU = \sum_{i \in L_T} \lambda_i \cdot C_i = \sum_{i \in L_T, SINR_i \geq \alpha} 2 \cdot w_{f_i} \cdot \log_2(M_i) \qquad (6)$$

This network utility definition is useful for elastic applications, such as those based on TCP, where the transmission rate can be adapted to the available resources. However, for inelastic applications with a fixed rate (e.g. VoIP), which require a minimum $SINR$ to guarantee a bit error rate, the only requirement is to establish the link with a $SINR$ higher than the predefined threshold $\alpha$. In this case, the capacity of link $i$ can be directly defined as $C_i$ = 1. And consequently, the network utility is defined as the number of valid links in the network:

$$NU = \sum_{i \in L_T} \lambda_i \cdot C_i = \sum_{i \in L_T, SINR_i \geq \alpha} 1 \qquad (7)$$

## 4. GAME THEORETIC AND LEARNING APPROACH

Regardless of the specific definition of the network utility $NU$ and the link capacity $C_i$, the generic problem of channel and power allocation can be modeled as a formal game defined as $\Gamma = \{N, \{S_i\}_{i \in N}, \{u_i\}_{i \in N}\}$, where $N$ is the finite set of players (the links) and $\{S_i\}_{i \in N}$ is the set of strategies $s_i = (p_i, f_i)$ (assignment of transmission power and frequency channel) related to player $i$. $S = \times_{i \in N} S_i$ is the strategy space and $u_i : S \to \Re$ is the set of utility functions that players associate with their strategies. For every player $i$ in the game $\Gamma$, $u_i$ is a function of $s_i$, the strategy selected by player *i*, and of *s₋ᵢ*, the current strategy profile of its opponents. Players will selfishly choose the actions that improve their utility functions considering the current strategies of the other players. Thus, the key issue is the choice of $u_i$, such as the individual actions of the players provide a good overall performance. Two properties are usually desirable: the game should have an equilibrium point and this point should maximize the network utility as defined in (3), (6) or (7). This equilibrium point, where no player has anything to gain by unilaterally deviating, is known as Nash Equilibrium (NE). Thus, a Nash equilibrium of a game Γ is a profile $s^* \in S$ of actions such that for every player $i \in N$ we have:

$$u_i(s_i^*, s_{-i}^*) \geq u_i(s_i, s_{-i}^*) \tag{8}$$

for all $s_i \in S$, where $s_i$ denotes any strategy of player $i$ and $s_{-i}^* \in S$ denotes the strategies of all players other than player $i$ in the profile *s\**.

Next we propose two different games adapted to this problem, with local and overall information respectively. In section 4.3, playing and timing rules for these games are defined. Then, additional power constraints are introduced for the local game in section 4.4. Since the convergence of the local game to a NE cannot be guaranteed, an alternative no-regret learning is described in section 4.5. The performance of the different proposals is analyzed in section 5.

### 4.1 Local game

Using only local information, the utility function for link *i* can be directly defined as the capacity of link *i*:

$$u_i(s_i, s_{-i}) = \begin{cases} -1 & \text{if } 0 < SINR_i < \alpha \\ C_i(s_i, s_{-i}) & \text{otherwise} \end{cases} \tag{9}$$

This expression can be performed locally since it can be directly calculated with the estimated *SINR* at the receiver. This *SINR* depends on the interference level for the different channels, the path loss between transmitter and receiver and the available transmission power levels. Those interference levels and the path loss can be estimated at the receiver and reported to the transmitter to calculate the utility.

The -1 value for *SINR* lower than the threshold but higher than 0 tries to introduce a degree of cooperation to compensate the inherent selfishness of this game: if a link cannot be established, it is better stop transmitting to reduce the interference of the remaining links. Namely, if the link is stopped, the transmission power is 0 and according to (9), $u_i(s_i, s_{-i}) = C_i(s_i, s_{-i}) = 0$. Therefore, this strategy is preferred to those with $u_i(s_i, s_{-i}) = -1$.

In this game, each link does not require any information about the strategies of the other players ($SINR_i$ can be measured without the specific knowledge of $s_{-i}$). On the contrary, existence and convergence to a pure NE cannot be assured.

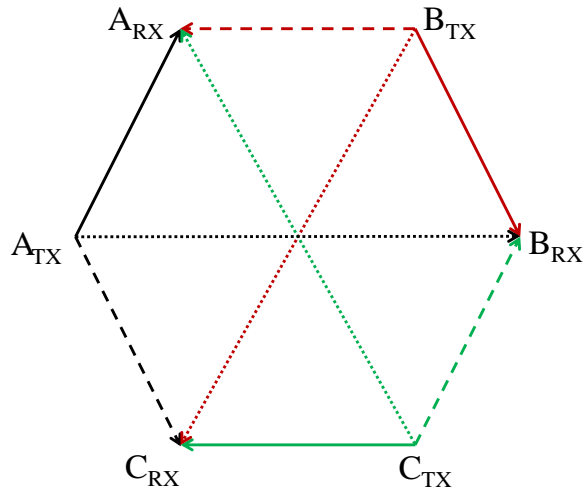

Figure 1. Network topology example of lack of a pure NE for the local game. Solid lines: links; dashed lines: high interference disrupting reception; dotted lines: low interference compatible with reception.

To show this possible lack of a pure NE, we use the following example. We assume a scenario with only three links (A, B and C) and two available channels, $f_1$ and $f_2$. The network topology shown in Fig. 1 is such that in both channels $f_1$ and $f_2$ only one link can simultaneously obtain a $SINR>\alpha$. In addition, these three conditions are also imposed by the network topology:

i) $SINR_A>\alpha$ only if B is not transmitting in the same channel, regardless of transmission power of C.

ii) $SINR_B>\alpha$ only if C is not transmitting in the same channel, regardless of transmission power of A.

iii) $SINR_C>\alpha$ only if A is not transmitting in the same channel, regardless of transmission power of B.

Given these three conditions, there is not a pure NE for this local game: if B is successfully transmitting in channel $f_1$ ($SINR_B>\alpha$) and C is successfully transmitting in channel $f_2$ ($SINR_C>\alpha$), link A will see that it can obtain $SINR_A>\alpha$ in channel $f_2$ (condition i). However, the activation of link A will make $SINR_C<\alpha$ (condition iii), so link C will stop transmitting and try to change to channel $f_1$, where it can obtain ($SINR_C>\alpha$). This will make $SINR_B<\alpha$ (condition ii), so now link B will stop transmitting and try to change to channel $f_2$. Again, this will make $SINR_A<\alpha$ (condition i) and link A will stop transmitting, returning to the initial condition, but with B and C transmitting in different channels. Due to the symmetry of the problem, it can be easily seen that no pure NE can be obtained in this scenario.

*4.2 Potential game*

Exact Potential Games (EPG) are a kind of games with very appropriate features to convergence purposes. An EPG is a game for which there exists a potential function $V(S)$, which is a function $V: S \to \Re$ with the property that:

$$\Delta u_i = u_i(s_i, s_{-i}) - u_i(s'_i, s_{-i}) = \Delta V = V(s_i, s_{-i}) - V(s'_i, s_{-i}), \quad \forall i \in N, \forall s_i, s'_i \in S_i \quad (10)$$

This definition implies that each player's individual interest is aligned with the groups' interest, since each change in the utility function of each player is directly reflected in the same change for the potential function.

If only one player acts at each time step (repeated sequential game) and the acting player maximizes (best response strategy) or at least improves (better response strategy) [4] its utility, given the most recent action of the other players, then the process will converge to a NE regardless of the order of play and the initial condition of the game. In addition, global maximizers of the potential function $V$ are NE, although they may be just a subset of all NE of the game. Therefore, defining a potential game with function $V$ equals to the network utility $NU$ can optimize the global capacity by individually playing the game.

A direct option is to define the utility function $u_i$ equal to the potential function, that is, the network capacity (identical interest games [35]):

$$u_i(s_i, s_{-i}) = \sum_{j \in L_T} \lambda_j \cdot C_j \tag{11}$$

In this game, each link $i$ requires global information about all the links in the network. Namely, the channel gain between any pair of nodes and the strategies $s_j = (p_j, f_j)$ of each link in the network. In addition, it requires for each strategy, to compute the capacity of all the links in the network. To reduce this computational complexity, a second possibility with the same potential function lies in defining the utility function according to (12):

$$u_i(s_i, s_{-i}) = \lambda_i \cdot C_i - \left( \sum_{j \in L_T, j \neq i} \lambda_{j,-i} \cdot C_{j,-i} - \sum_{j \in L_T, j \neq i} \lambda_j \cdot C_j \right) \tag{12}$$

where $\lambda_{j,-i}$ and $C_{j,-i}$ represent the same as $\lambda_j$ and $C_j$ if link $i$ were not active.

The expression reflects the intention to maximize the player's own utility, but subtracting the potentially negative effect over other players. To this purpose, the second term expresses the increase in the capacities of the remaining links of $i$ if this link were not active, and it is calculated as the difference between these capacities without and with the active link $i$. These $C_{j,-i}$ and $C_j$ values are only different, and therefore, only have to be calculated, if both $i$ and $j$ are transmitting in the same channel. Moreover, the term $C_{j,-i}$ does not have to be recomputed for each strategy, since it is independent of that. Thus, this utility function requires less computational load than the one of the previous identical interest game. It can be easily checked that the network capacity is again a potential function of this game, which according to (10) is an EPG. Both games have the same potential function and require the same network information. Thus, in the performance evaluation section we will focus in this second option, since it has lower computational complexity.

### 4.3 Playing and timing rules

Both local and potential games are played as a myopic repeated game. A repeated game is a sequence of stage games where each stage game or step is the same normal form game. The myopic term reflects that each player takes its decision according to the observation of the most recent scenario where it is playing, instead of considering past actions or future expectations [3]. Thus, complex multi-stage strategies are not possible. However, simpler myopic strategies, such as the best or better response dynamics, can be used. As briefly introduced in section 4.2, a better response is a playing rule that decides to change to a new strategy that at least improves the current utility. A best response corresponds to a playing rule that decides to change to the strategy that provides the optimum utility given the current opponents' profile.

In addition to the playing rule, a repeated game is characterized by the specific timing followed by the players, that is, the playing order. Two different criteria to determine this order have been considered in this work:

- **Round Robin scheduling**: at each step only a single player plays. This implies that this player will not update its strategy again until all the remaining players have played. This scheduling guarantees the convergence for the potential game, since as stated above, any potential game in which players take actions sequentially under a best or better response strategy converges to a pure NE. However, a central coordinator is required to set the playing order, which is not realistic in the considered scenario.

- **Asynchronous scheduling**: at each step all players have a probability $1/N$ (being $N$ the number of players) of updating their strategy. On average, only one player actually updates, however, in practice, several players can simultaneously change their strategy at the same step. Although it does not fit the single player updating rule, in [14] and [16] the convergence of this approach for a potential game is verified by simulations. This scheduling models the interactions among the players more realistically, since it shows that converge is possible in a distributed scenario and, accordingly, this approach is also considered in the presented work.

## *4.4 Power constraints*

With the network utility definition of (3), strategies with different *SINR* provide different utilities. However, when a more realistic scenario is considered, the definition of the network utility according to (6) and (7) leads to a playing rule that evaluates different strategies with the same utility regardless of the *SINR*. For example, when maximizing the probability of establishing a link, different channels may guarantee a *SINR* over the threshold and in any of them, all the transmitted powers that fulfill this *SINR* provide the same utility. An analogous effect occurs when a modulation is selected and several available transmission powers provide the same utility. This equivalence in utilities can make the game converge to a suboptimum global performance, since better strategies from a global point of view (lower transmission power to reduce interference) will have no preference. That is, when a valid *SINR* is possible, players try to maximize their own capacity without considering the influence on the others. An alternative approach to increase cooperation is to reflect this compromise in the utility function. The expression in (13) compensates the utility with a factor dependent on the transmission power. Thus, strategies that provide the same capacity can be differentiated by the selected transmission power ($p_i$). The designed factor $w_{f_i} \times (1-p_i/P_{max})$ ensures the predominance of lower transmission powers only in the conflicting cases with the same capacity, without disrupting the original decision rule according to capacity.

$$u_i(s_i, s_{-i}) = \begin{cases} -1 & \text{if } 0 < SINR_i < \alpha \\ C_i(s_i, s_{-i}) + w_{f_i}\left(1 - \dfrac{p_i}{P_{max}}\right) & \text{otherwise} \end{cases} \quad (13)$$

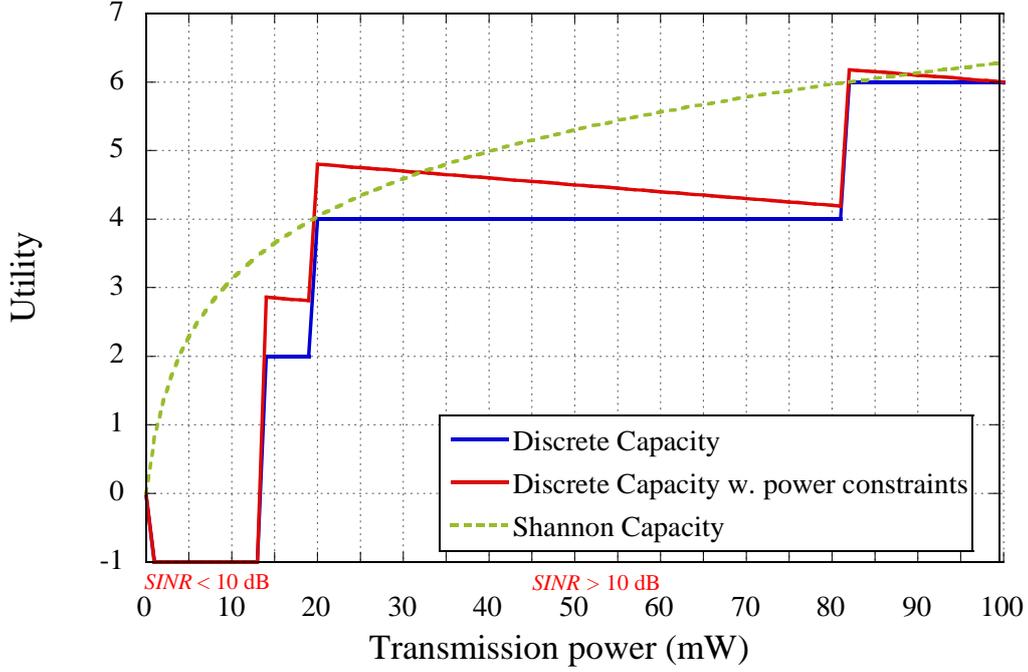

Figure 2. Example of utility function with power constraints in a specific channel. $s_i = (f_i, p_i)$, $f_i = f$, $p_i = \{0,...,P_{max}\}$, $P_{max} = 100$ mW. Fixed level of sensed interference in $f$ according to $s_{-i}$.

Fig. 2 shows an example of the utility functions versus the transmission power with – Eq. (13) – and without – Eq. (9) – this power correction for the discrete capacity $C_i = 2 \cdot w_{f_i} \cdot \log_2(M_i)$ with $M_i$ a function of the *SINR*. This *SINR* results from a fixed sensed interference in the selected reference channel and thus it is only dependent on the transmission power. The Shannon capacity is also included as a reference. As can be seen in the figure and assuming $w_{f_i}$ is normalized to 1 unit, discrete capacity can take the values 2 ($M_i = 2$), 4 ($M_i = 4$) and 6 ($M_i = 8$) depending on the transmission power (and therefore the obtained *SINR*). For example, for $C_i = 4$, the utility at low transmission powers is close to 5, since $p_i/P_{max}$ tends to 0 and at high transmission powers is close to 4, since $p_i/P_{max}$ tends to 1. In any case, the correction factor will never disrupt the original decision according to capacity, since the highest utility with $M_i = 2$ will be always lower than 3 and the lowest utility with $M_i = 8$ will be always higher than 6.

*4.5 No-regret learning*

As explained in section 4.1, the designed local games do not have the desired properties to guarantee the convergence to a NE (as a potential game ensures). In order to analyze an alternative approach to guide the reconfiguration process with the same local information, we implement the adaptation of the decision rules according to a no-external-regret learning algorithm. This learning method provides a valuable mechanism to lead the repeated game to a new set of equilibria, the coarse correlated equilibria introduced in section 2, which may provide global utilities even beyond the Nash solution [10].

A coarse correlated equilibrium is a probability distribution $\pi = \left(\pi\left(s^{(1)}\right),...,\pi\left(s^{(|S|)}\right)\right)$ over the set of joint actions $S = \left(s^{(1)},...,s^{(|S|)}\right)$, with $\pi\left(s^{(n)}\right)$ the probability of observing the action profile $s^{(n)}$ that satisfies:

$$\forall i \in N, \forall s'_i \in S_i, \quad \sum_{s_{-i} \in S_{-i}} u_i\left(s'_i, s_{-i}\right)\pi_{-i}\left(s_{-i}\right) \leq \sum_{s \in S} u_i\left(s_i, s_{-i}\right)\pi\left(s_i, s_{-i}\right) \quad (14)$$

where $\pi_{-i}\left(s_{-i}\right)$ represents the marginal distribution of all players other than player $i$, and is given by:

$$\pi_{-i}\left(s_{-i}\right) = \sum_{s_i \in S_i} \pi\left(s_i, s_{-i}\right) \quad (15)$$

The learning process in the repeated game involves playing a mixed strategy to infer the expected utilities of all available strategies against any opposing profile. In the previously defined game Γ, $Q_i$ denotes the set of mixed strategies for player $i$ over its set $S_i$. According to $Q_i$, $q_i$ is the probability of choosing the strategy $s_i \in S_i$ in each stage of the repeated game.

Generally speaking, the regret $\rho_i$ a player feels when playing a strategy $s_k$ instead of a strategy $s_i$ in time $t$ is defined as the difference in the obtained utilities: $\rho_i\left(s_i, s_k^t \mid s_{-i}^t\right) = u_i\left(s_i, s_{-i}^t\right) - u_i\left(s_k^t, s_{-i}^t\right)$. A no-regret learning algorithm $A_i: H_i \to Q_i$ tries to sequentially redefine the probabilities according to the history $\left(h_i^t\right)$ observed during the repeated actions of the game $q_i^{t+1} = A_i\left(h_i^t\right)$, with the goal of minimizing the expected regret. $A_i$ is said to exhibit no-external-regret iff, by definition, the cumulative average regret tends to zero in the limit ($T \to \infty$) w.r.t. all possible opposing sequences of play $\left\{s_{-i}^t\right\}$ [13].

We have evaluated two no-external-regret learning algorithms. One of them is based in the exponential updating approach defined by Freund and Schapire – FS – [31], and the other one follows the Hart&Mas-Colell – HM – proposal [30]. Each user updates the probabilities of its mixed strategy $Q_i$, well according to the perceived utility (FS), either the experienced regret (HM): In any case, the algorithm tries to increase the probability of playing those strategies that provide higher expected utilities (lower regret).

*Algorithm 1 – FS($\beta$)*: The cumulative utility obtained by user $i$ through time $t$ by choosing strategy $s_i$ is given by $U_i^t(s_i) = \sum_{x=1}^{t} u_i\left(s_i, s_{-i}^x\right)$. For $\beta > 0$, the probability assigned to strategy $s_i$ at time $t+1$ is given by:

$$q_i^{t+1}(s_i) = \frac{(1+\beta)^{U_i^t(s_i)}}{\sum_{s'_i \in S_i}(1+\beta)^{U_i^t(s'_i)}} \tag{16}$$

*Algorithm 2* – HM: The cumulative regret felt by player *i* for not having played strategy $s_i$ through time *t* is given by $R_i^t(s_i) = \sum_{x=1}^{t} \rho_i(s_i, s_i^x | s_{-i}^x)$, $\rho_i(s_i, s_i^x | s_{-i}^x) = u_i(s_i, s_{-i}^x) - u_i(s_i^x, s_{-i}^x)$. The update rule, in this case, is given by:

$$q_i^{t+1}(s_i) = \frac{\left[R_i^t(s_i)\right]^+}{\sum_{s'_i \in S_i}\left[R_i^t(s'_i)\right]^+} \quad (17)$$

where $X^+ = \max\{X, 0\}$.

In both cases, we assume informed players [13]. An informed player observes the strategy it plays and the vector of utilities it would have obtained if it had played any of all its possible strategies. Thus, at each step *t*, a player randomly chooses a strategy $s_i$ according to the current probabilities, and then it updates all the probabilities for all the possible strategies according to the utilities it would obtain in this scenario.

The assumption of informed players is the same internal process that each player makes in the local game to estimate the utility of the set of possible strategies in order to choose one of them with a best or better response strategy. The utility definition for any strategy is a function of the estimated *SINR* at the receiver, and can be locally estimated as described in section 4.1 to perform the update.

## 5. RESULTS

The analysis and evaluation of a game model should cover two different aspects: on the one hand, the existence of some equilibrium points for the proposed game. On the other hand, the quality of these equilibria measured as the ratio between the network utility obtained in the equilibrium and the optimum network utility.

With regard to the first issue, the convergence of the different proposed games and learning approaches has been analyzed in section 4. First, it has been demonstrated with an example that the existence of a pure Nash equilibrium cannot be guaranteed in a generic scenario for the proposed local game. Given the well-known properties of the potential games [3], the existence and convergence to a Nash equilibrium is guaranteed for the considered potential game. Finally, in section 4.4 it is described that no-external-regret learning algorithms converge to the set of coarse correlated equilibria as it is shown in [30].

Regarding the quality of the equilibria, due to the complexity of the channel and power allocation problem to maximize network capacity, the proposed game learning approach have been evaluated by simulations and compared to a heuristic centralized genetic algorithm which estimates the optimum performance of the network.

Several scenarios have been studied varying the values of the main simulation parameters (topology size, number of nodes, noise power, available frequency channels, power levels in the transmitter). The relative differences among the strategies hold in any case, making the proposals scalable and applicable to diverse situations. For clarity, this paper presents results only in the scenario described below. The specific presented results are a sample of the correctness of the proposed games. Since the relative performance, and therefore the obtained conclusions are maintained, the latter can be extensible to different settings.

The network consists of 200 nodes deployed in a square area of length 2.4 km with different random topologies. Several numbers of links ($N$ players of the game) ranging from 50 to 400 are considered in the evaluation. The total number of non-interfering channels is set to $F = 10$. To model the presence of other non-cognitive users, some channels are not sensed as available for the nodes in the network. The set of available channels varies depending on the location. Specifically, at regions of 100x100m, a different random and variable in size subset from 3 to 8 channels is sensed as available for nodes in that region. $P_{max}$ is set to 20 dBm. Transmission power is quantized with $Q = 16$ levels. The path loss index is $\gamma = 4$ and for simplicity, the bandwidth of each channel, $w_{f_i}$, is the same and normalized to 1 unit. The *SINR* threshold is set to 10 dB. $P_N$ is set to -85.9 dBm, which ensures a *SINR* of 10 dB at 250 m, then fixing the maximum transmission range in 250 m.

## 5.1 Local vs. potential game

In this subsection we analyze the behavior of the local and potential games proposed in sections 4.1 and 4.2. As a first step, the power constraints defined in section 4.4 are not included to perform a fair comparison between both games only differentiated in the selfish or common interests covered by the utility functions. Three settings have been evaluated. They differ in the utility function used by players to select their strategy and the definition of users' capacity.

a) Continuous capacity according to Shannon-Hartley theorem ($C_i = w_{f_i} \cdot \log_2(1 + SINR_i)$) with $\alpha = 0$ (CC-no$\alpha$) and $\alpha = 10$ (CC-$\alpha$). In this scenario, users' goal is to maximize a continuous capacity: individual in the local game or aggregate in the potential one. The *SINR* constraint is not considered in the utility function of CC-no$\alpha$, that is, users play continuously transmitting to improve capacity as presented in previous related works [14]-[17]. Hence, we include this initial assumption as a reference. However, to make results comparable, we also measure the actual obtained *SINR* and the number of valid links ($SINR_i > \alpha = 10$), which can actually evaluate the real performance of this game. In contrast, in CC-$\alpha$ the *SINR* constraint is considered in the decision process. Users do not continuously transmit, but remain silent if the *SINR* threshold cannot be satisfied.

b) DC-$\alpha$: Discrete capacity, with 8 available modulation levels. $C_i = 2 \cdot w_{f_i} \cdot \log_2(M_i)$, $2 \leq M_i \leq 256$, and $M_i$ a function of the *SINR*, as defined in section 3. In this scenario, users' goal is to maximize a discrete capacity, individual (local game) or aggregate (potential game). This is an extension of CC-$\alpha$ with discrete capacity.

c) BC-$\alpha$: Binary capacity (1 or 0 depending on the transmission success) according to (7) with $\alpha$ = 10. This scenario is equivalent to DC-$\alpha$, but the users' goal is not to maximize the capacity (bps), but the opportunity to establish valid links (transmission with *SINR* > $\alpha$). In this case, the local game interest is to activate the own link, whereas the potential game focuses on establishing the maximum number of links in the network.

Different instances of the games (1000 plays for each value) have been simulated and averaged in order to obtain mean values of the performance of every strategy. In each instance, the position of the nodes, the selected links and the initial conditions – strategies ($p_i, f_i$) – are randomly generated. The game is played until a pure Nash equilibrium is found or until a predefined maximum number of iterations is reached. This number has been empirically selected to ensure that the NE is not going to be reached later (20000 steps in the presented results).

Since it has been evaluated that there are not significant differences with regard to the results obtained with a better response strategy and it exhibits a faster convergence, players follow a best response criterion to select their strategy: the user evaluates strategies channel by channel from the lowest to the highest transmission power at each channel. An additional restriction is included for the potential game: the self interest ($C_i$) is prioritized to choose among strategies with the same utility ($s_i$ and $s_i^*$) That is, $\left[ u_i(s_i, s_{-i}) = u_i(s_i^*, s_{-i}^*) \right] \wedge \left[ C_i > C_i^* \right] \Rightarrow s_i$ is better than $s_i^*$.

As described in section 4, there are two possible criteria to determine the players' order: round robin and asynchronous scheduling. We have verified a similar performance in both cases (almost equal *NU*) mainly differing in the convergence speed (asynchronous is slower). Since this last one represents more accurately a distributed scenario, the presented results refer to this asynchronous approach hereafter.

Fig. 3 shows the obtained network utility with continuous capacity without (CC-no$\alpha$) and with *SINR* threshold (CC-$\alpha$) with the proposed games. In the first case, it can be seen that the obtained capacity with the local game is clearly lower than with the potential game, especially as the density of links grows. For example, with 400 links, the network utility with the local game is around an 80% of the potential game. These network utilities are obtained assuming each link can be established despite of the obtained *SINR*. When no restriction is included to establish a link (*SINR* over threshold), all users in the local game selfishly try to maximize its own capacity leading to maximum transmission power. Therefore, a higher level of interference in the network reduces the aggregated capacity regarding the situation of the potential game. In this latter, power control is activated due to the trade-off expressed in the utility function that compensates the negative effect over the other players.

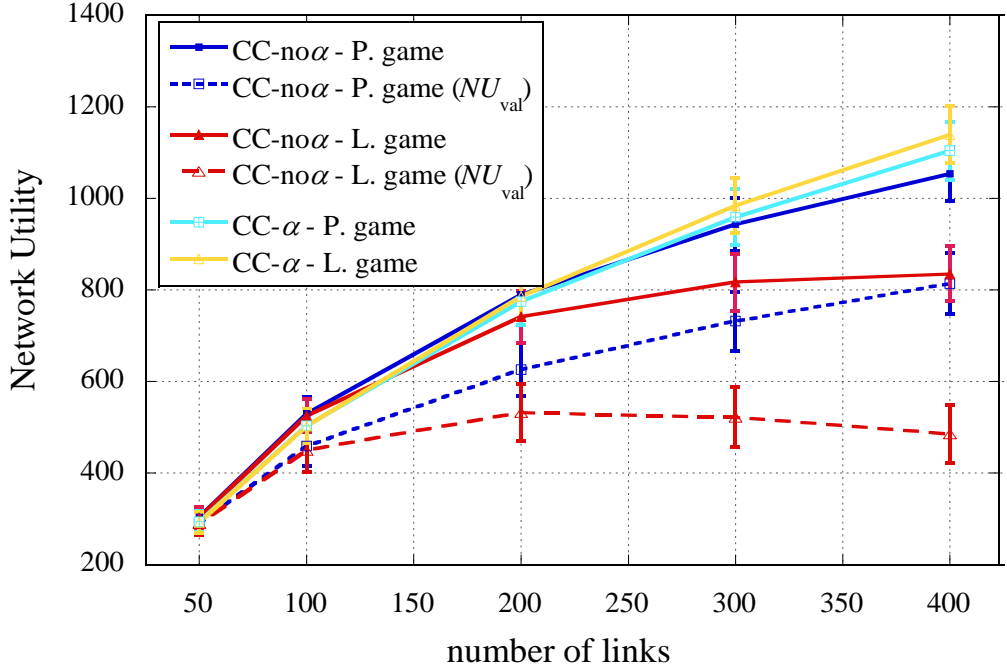

Figure 3. Network utility. Mean value and standard deviation (bars). $NU_{val}$: $NU$ evaluated adding only valid links ($SINR_i > \alpha$).

In addition, if we only take into account the capacity of valid links ($NU_{val}$), i.e., those with $SINR > 10$, the actual utility is clearly lower in both cases. This happens because of the lack of an admission control that ensures the establishment of only feasible links. In this sense, the inclusion of the $SINR$ threshold provides a more realistic framework. In this scenario, for the strategy CC-$\alpha$, the local game obtains similar capacity to the potential game. The $SINR$ restriction tries to solve the admission control problem. In the potential game, only feasible links are activated and, in addition, power control limits the transmission power to allocate more links if possible. In the case of the local game, users still focus only on maximizing their own capacity transmitting at maximum power. However, this is only done when the link is feasible due to the unselfishness provided by the -1 value for $SINR < \alpha$ in (9). This enables to approach, even outperform, the capacity of the potential game (Fig. 3). In contrast, the lack of an actual power control that limits the transmission power slightly reduces the number of valid links, as it is observed in Fig. 4.

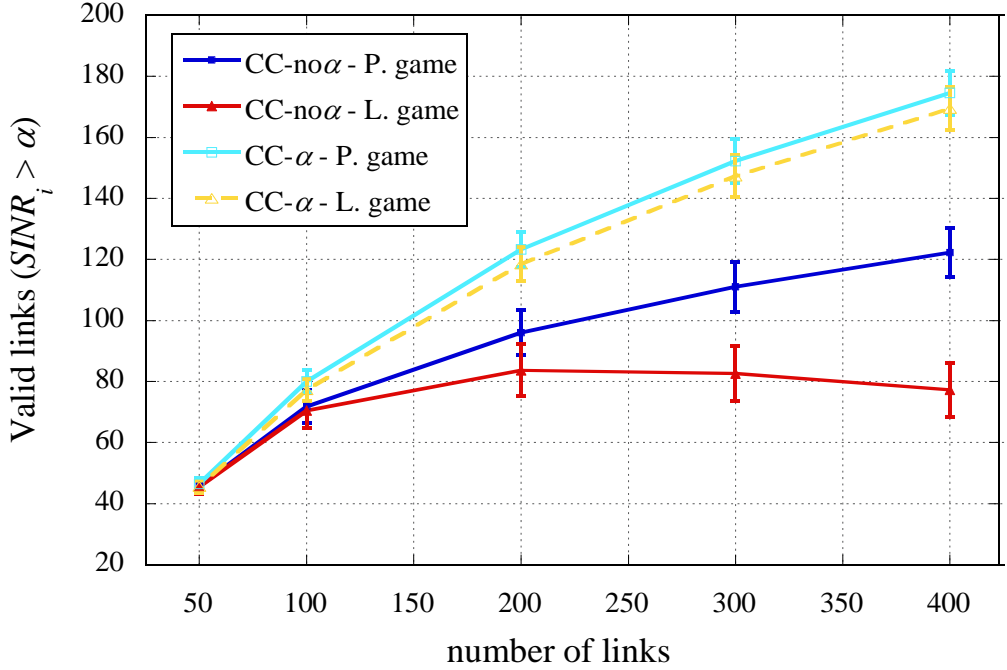

Figure 4. Number of valid links ($SINR > \alpha$). Mean value and standard deviation (bars).

Concluding the necessity of an admission control mechanism, as a next step, we focus on the more realistic strategies DC-$\alpha$ and BC-$\alpha$. They both try to optimize different network utilities (capacity in bps the first one and number of valid links the second one) that represent the two conflicting goals described above (maximizing capacity, admission control). Anyway, for comparison purposes, we will show both metrics for both games. To evaluate the goodness of the proposals, results are compared to the solution provided by a Genetic Algorithm (GA) [20], which gives an estimation of the optimal performance of the network. The solution results from the average of 50 random topologies with the parameters shown in Table 1. The GA has been used to maximize the network utility *NU*, using information of all the links in the network. According to the DC-$\alpha$ and BC-$\alpha$ strategies, two definitions of *NU* have been considered, as defined in section 3: maximizing discrete network capacity (6) and maximizing the number of valid links (7). Although this algorithm does not guarantee the optimum value, it provides a valuable reference for the performance of the games.

TABLE 1. PARAMETERS - GENETIC ALGORITHM SIMULATIONS [20]

| Parameter | Value |
|---|---|
| Population size | 1000 |
| Maximum generations | 20000 |
| Number of variables | $2 \times N$ ($p_i, f_i$) |
| Replace proportion | 0.9 |
| Selection method | Tournament with replacement (size 500) |
| Crossover prob. / method | 0.9 / Simulated binary crossover (genewise swap probability 0.5, polynomial order 10) |
| Mutation prob. / method | 0.1 / Selective |
| Constraint-handling / Constraints | Tournament / $u_i = 0$ ($SINR_i < \alpha$) $\Rightarrow p_i = 0$ |

Fig. 5 and Fig. 6 respectively show the capacity and the number of active links for both games BC-$\alpha$ and DC-$\alpha$. The potential game provides a performance similar to the centralized GA in both cases, which agree with the principle of maximizing a potential function equals to the network utility. This good performance comes at the expense of each link requiring overall information. However, as in the strategy CC-$\alpha$, in both cases, the local game provides a similar performance to the potential game with much less information, which makes it a more suitable solution. Moreover, with both BC-$\alpha$ and DC-$\alpha$ strategies, players in the local game no longer try to transmit at maximum power at any case, since the same utility can be obtained with lower transmission powers (players evaluate strategies channel by channel from the lowest to the highest transmission power at each channel, as described before). Thus, for the DC-$\alpha$ strategy, the number of active links with the local game is even higher than with the potential game. In addition, in the BC-$\alpha$ strategy, both local and potential game perform equal with regard to the number of active links (the aim of the game), but the local game even achieves a higher capacity, providing better global performance.

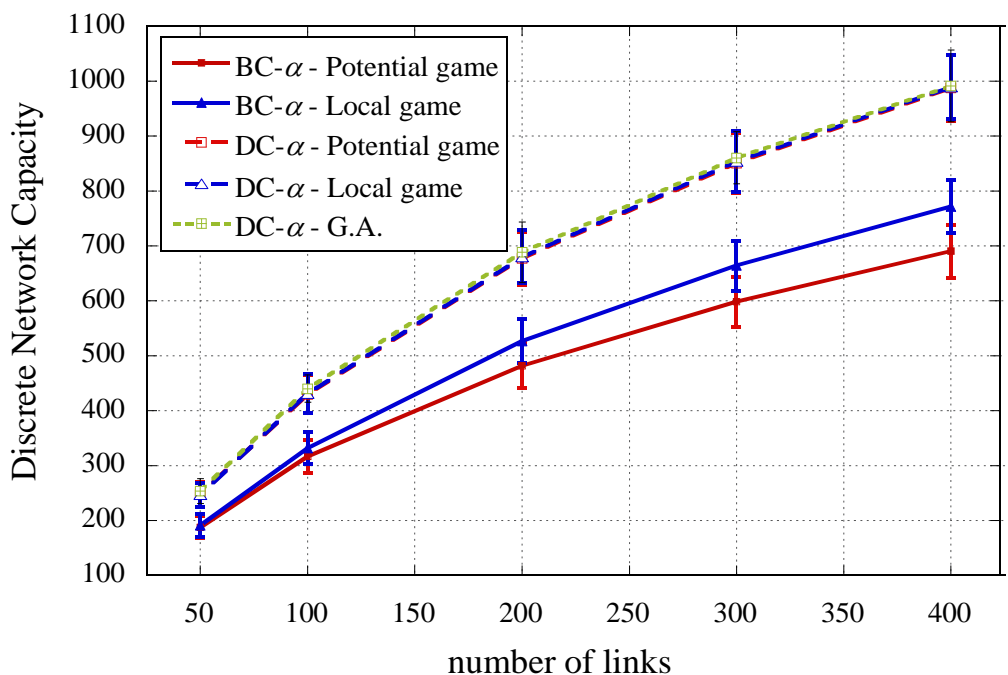

Figure 5.  Discrete network capacity. Mean value and standard deviation (bars). Strategies BC-$\alpha$ and DC-$\alpha$ compared to a centralized genetic algorithm (optimization of *NU* in DC-$\alpha$).

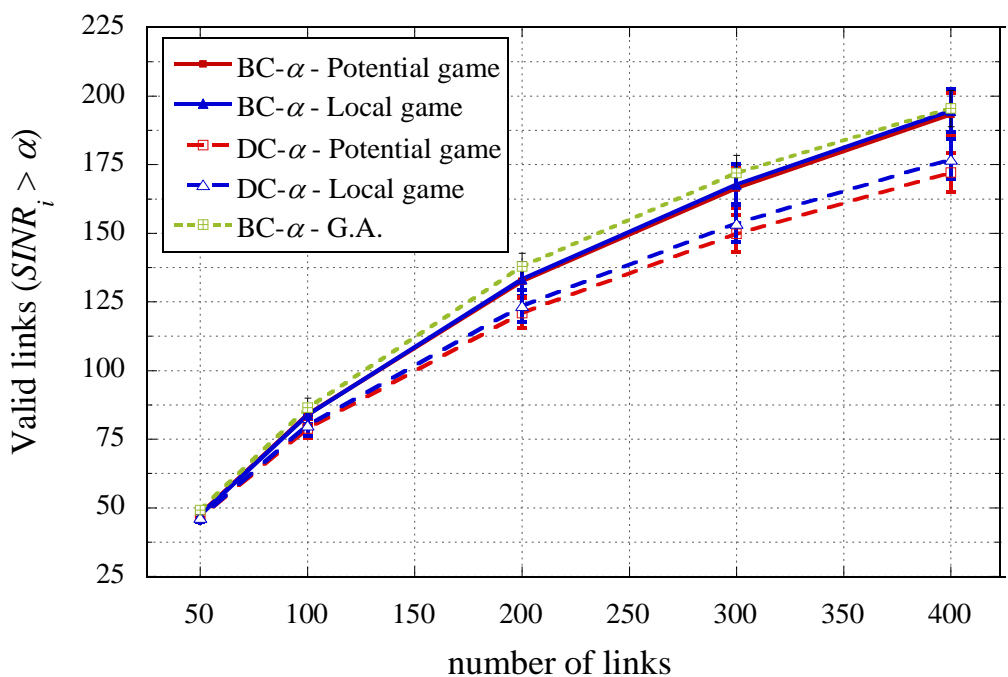

Figure 6.  Number of valid links. Mean value and standard deviation (bars). Strategies BC-$\alpha$ and DC-$\alpha$ compared to a centralized genetic algorithm (optimization of *NU* in BC-$\alpha$).

In addition, although convergence to a pure NE cannot be guaranteed for the local game, the convergence probability obtained in the simulations is very high for both BC-$\alpha$ and DC-$\alpha$, being always higher than 99% in all the performed simulations (the potential games always converge, as explained in section 4.2). Furthermore, despite the higher convergence speed of the potential game (15 − 20% higher than the local approach – Fig. 7), the reduction of complexity and information required by the local game make the latter a preferred scalable solution. In Fig. 7 each iteration represents a player action, i.e., a link action. Therefore, it can be seen that the number of iterations per link remains approximately constant as the number of links grows (about 100 − 150 iterations per link).

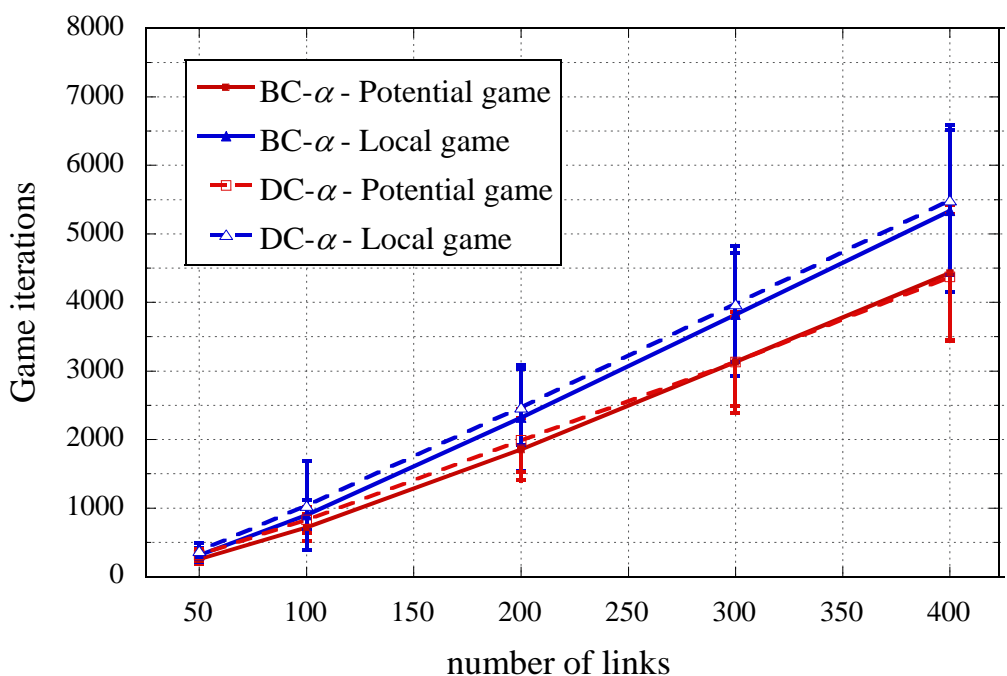

Figure 7.  Number of game iterations. Mean value and standard deviation (bars). Strategies BC-$\alpha$ and DC-$\alpha$.

## 5.2 No-regret learning for local game with power constraints

As stated in section 4.1, the existence of a pure NE cannot be assured for the local game. In addition, as explained in section 4.4, the performance of the local game could be improved if each player always selected, for the same utility, the strategy with a lower transmission power (if the link is feasible, maximize capacity without wasting power). This would reduce the interference created to the remaining players without decreasing its own utility, but improving the aggregate one. To that purpose, the utility function can be defined according to (13). When playing both local games BC-$\alpha$ and DC-$\alpha$ with this utility function under a best response strategy, it has been observed that they do not converge to a NE.

For these reasons, we study the performance of this local game with power constraints using the no-regret learning approach described in section 4.5. Specifically, we have used two no-external-regret algorithms (FS with $\beta = 0.1$ – Eq (16) and HM – Eq (17)), which ensure the convergence to the set of coarse correlated equilibria. Similarly to the results presented in the previous section, different instances of the learning algorithms have been simulated and averaged in order to obtain the mean values shown in the following figures. In this scenario, the evaluated games with power constraints are denoted as DCP-$\alpha$ and BCP-$\alpha$, corresponding to their equivalent DC-$\alpha$ and BC-$\alpha$ without power constraints.

Fig. 8 and Fig. 9 respectively show the capacity and the number of active links for both games BCP-$\alpha$ and DCP-$\alpha$, compared to BC-$\alpha$ and DC-$\alpha$. These latter games with utilities without power constraints provide a reference solution (the reached NE) not achievable following the best response criterion in BCP-$\alpha$ and DCP-$\alpha$. It can be seen that both learning algorithms provide a very similar performance, both improving the results obtained with the local game: DCP-$\alpha$ improves both the network capacity and the number of valid links regarding DC-$\alpha$. In BCP-$\alpha$, the network capacity (which is not the objective of the game) decreases respect to BC-$\alpha$, since players use lower transmission powers, only trying to fulfill their *SINR* threshold $\alpha$. However, the increase in the number of valid links (the actual objective of the game) is higher, since the reduction in the overall generated interference enables more links to be active.

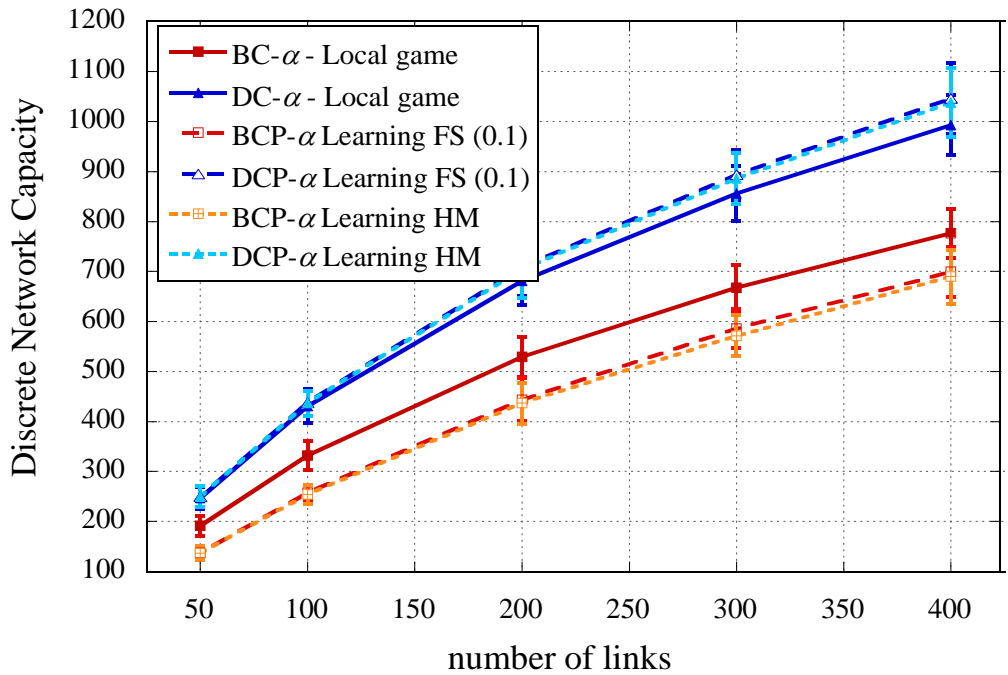

Figure 8. Discrete network capacity. Mean value and standard deviation (bars). Strategies BC-$\alpha$ and DC-$\alpha$ with the local game and the learning algorithms (FS, $\beta = 0.1$ and HM).

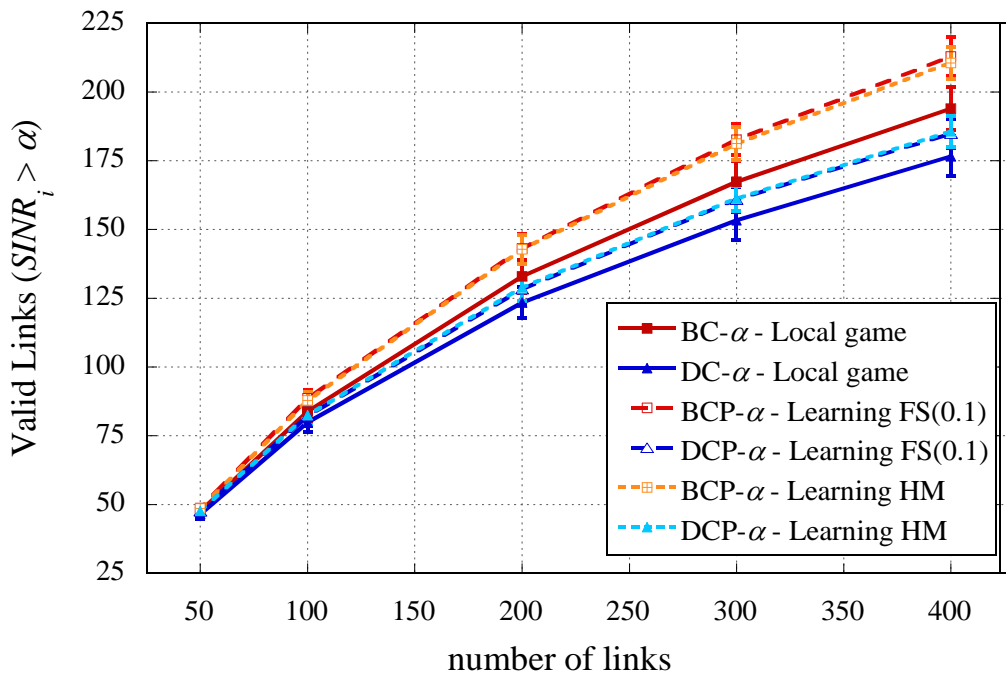

Figure 9. Number of valid links. Mean value and standard deviation (bars). Strategies BC-$\alpha$ and DC-$\alpha$ with the local game and the learning algorithms (FS, $\beta = 0.1$ and HM).

If we compare these results with those of Fig. 5 and Fig. 6, we can see that the learning algorithms even outperform the genetic algorithm. Although this result may seem odd, it is completely possible since genetic algorithms are not guaranteed to find the global optimum solution, but an approximate solution close to the actual global optimum. For any evolutionary algorithm, such as genetic algorithms, a solution is better only in comparison to other, already known solutions. These algorithms actually have no way to test whether a solution is optimal. In fact, for this reason, evolutionary algorithms are especially useful on problems where it is difficult or impossible to verify the optimality. Thus, a genetic algorithm never knows for certain when to stop, rather than a predefined total number of iterations or a given number of iterations without any change. Nevertheless and due to the random nature of the genetic algorithms, it cannot be assured that further iterations will not lead to a solution closer to the global optimum [36].

As previously stated, no-external-regret algorithms do not guarantee the convergence to a pure NE, but to the set of coarse correlated equilibria, where players can act with pure or mixed strategies. As an example, Fig. 10 shows the evolution of the probabilities $q_i$ of choosing each strategy $s_i$ for two players (solid lines for Link 1 and dashed lines for Link 2) in a scenario with 50 links using the FS algorithm for the discrete capacity. It can be seen that Link 1 converges towards a pure strategy (solid), whereas Link 2 does it towards a mixed one (dashed), where three strategies converge to a probability higher than 0.

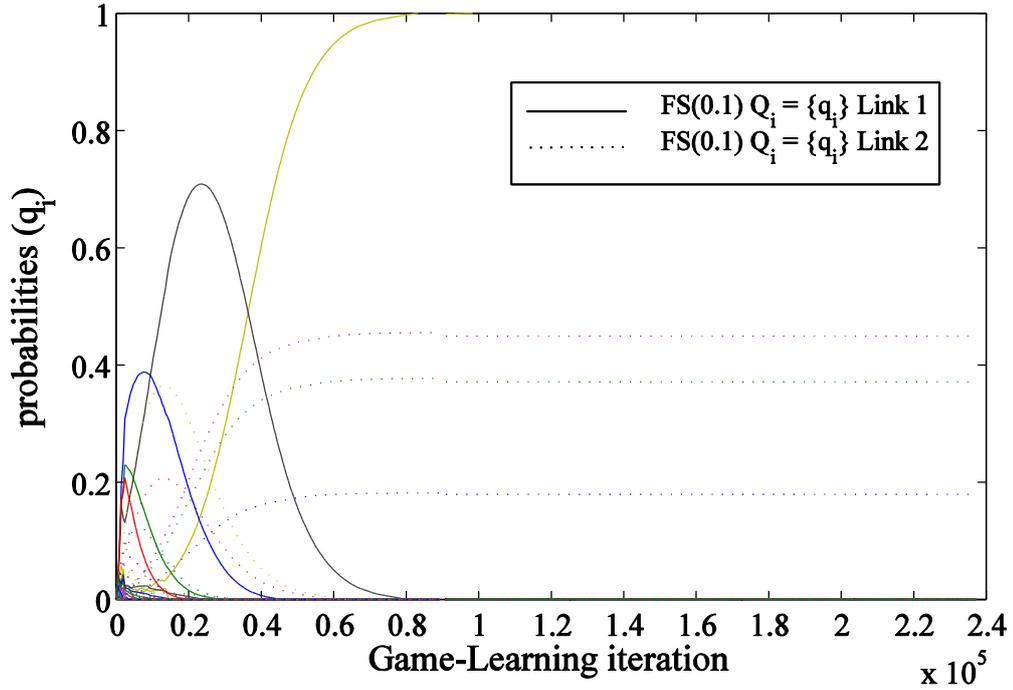

Figure 10. Example of the evolution of a mixed strategy solution ($q_i \in Q_i$) for two specific links in DCP-$\alpha$. Learning algorithm FS with $\beta = 0.1$. Scenario of 50 links. The solid lines correspond to the probabilities of the strategies of link 1, while the dashed lines correspond to the probabilities of the strategies of link 2.

In Fig. 11, it is shown an example of the evolution of the network utility in DCP-$\alpha$ using the two no-regret algorithms in a specific scenario with 200 links. Since players do not have to converge to a pure strategy, the specific network utility at each step does not remain constant, but slightly varies around the mean network utility, given by the joint probability distribution of the players' actions.

As can be seen, the convergence is slower than with the repeated game under a best response strategy (Fig. 7). However, the capacity of the no-regret learning algorithms to converge to a broad concept of equilibrium makes it an interesting solution for scenarios where the existence of NE cannot be guaranteed, such as the proposed local game with power constraints. In fact, the presented results suggest that although a pure NE cannot be achieved, the alternate selection of strategies by users in the mixed solution (for example, similar to a channel selection based on frequency hopping) enables to obtain a better average aggregate performance. This shows an interesting approach to ensure a viable joint power and channel allocation in a multihop cognitive network where a fixed optimum solution could be unrealizable from a practical perspective.

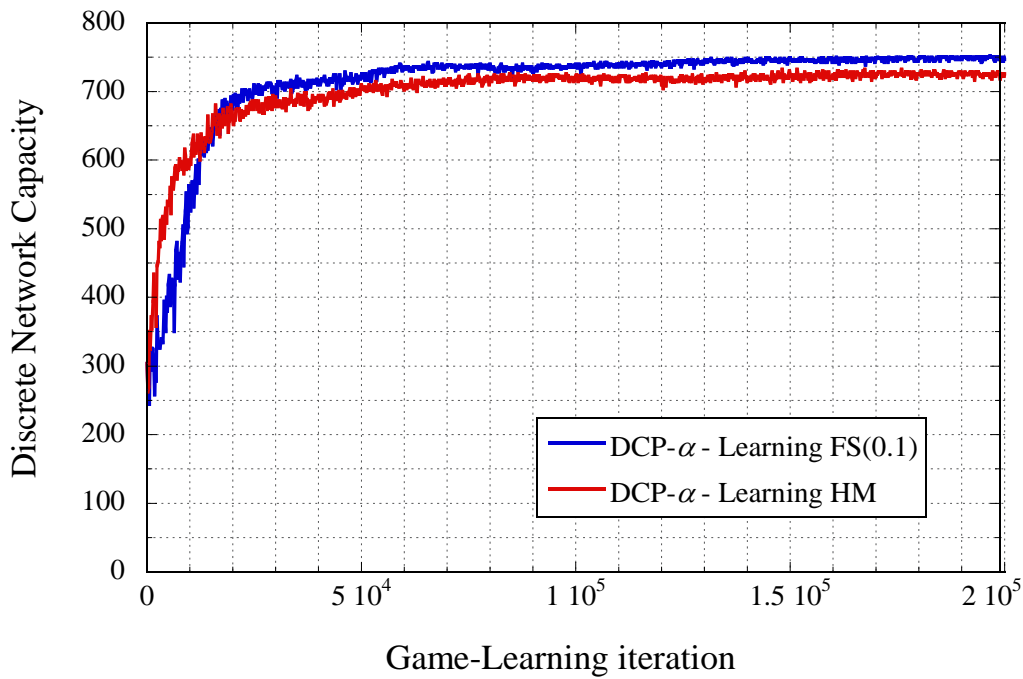

Figure 11. Evolution of discrete network capacity in DCP-$\alpha$. Scenario of 200 links. Comparison of learning algorithms FS, $\beta = 0.1$ and HM.

## 6. CONCLUSIONS

A game theoretic framework to perform distributed joint channel and power allocation in cognitive radio networks has been proposed and evaluated. The problem has been analyzed under the physical interference model, including in the decision process an application-dependent definition of the network utility and the successful transmission constraint (*SINR* over threshold).

Simulation results have shown the capability of the proposed game with local information and a low grade of cooperation to provide a performance similar to a less scalable potential game. Moreover, it ensures a global performance close to that obtained with a centralized genetic algorithm. A refined version of the defined utilities to increase cooperation via power correction has also been evaluated under a no-external-regret learning approach. This point of view overcomes the convergence limitations of the local game, providing even better global performance.

This opens an interesting perspective to develop realistic protocols based on the modeled interactions capable of performing efficient opportunistic spectrum access. In this sense, different challenges can be addressed by taking this line of research. Specifically, the learning approach can extend the game theory framework by introducing more flexibility in the scenario. As suggested by the provided results, learning can tackle the problem of unknown and variant opponents to converge to more generalized equilibrium solutions. Thus, as a natural evolution of this paper, it can be possible to model the interaction among users with more complex and differentiated utilities (e.g., links demanding different rates or QoS with uneven *SINR* requirements). Furthermore, the multihop routing issue needs to be incorporated as a generalization of the game learning approach to efficiently allocate end-to-end resources in a cognitive network.

## ACKNOWLEDGMENT

This work has been supported by the Spanish Government through the grant TEC2011-23037 from the Ministerio de Ciencia e Innovación (MICINN) and Fondos Europeos de Desarrollo Regional (FEDER).

# Biographies

**José Ramón Gállego**

José Ramón Gállego was born in Zaragoza (Spain) on 1978. He received the Engineer of Telecommunications MS and Ph.D. degrees from the Universidad de Zaragoza, Spain, in 2001 and 2007, respectively. In 2002, he joined the Centro Politécnico Superior, Universidad de Zaragoza, where he is currently an Associate Professor. He is member of the Aragón Institute of Engineering Research (I3A). His professional research activity lies in the field of wireless communications, with emphasis on radio resource management and mobility support, in 3G/4G, ad-hoc and cognitive networks.

**María Canales**

María Canales was born in Zaragoza (Spain) on 1978. She received the Engineer of Telecommunications MS and Ph.D. degrees from the Universidad de Zaragoza, Spain, in 2001 and 2007, respectively. In 2002, she joined the Centro Politécnico Superior, Universidad de Zaragoza, where she is currently an Associate Professor. She is member of the Aragón Institute of Engineering Research (I3A). Her professional research activity lies in the field of wireless communications, with emphasis on radio resource management and mobility support, in 3G/4G, ad-hoc and cognitive networks.

**Jorge Ortín**

Jorge Ortín was born in Zaragoza, Spain, in 1981. He received the Engineer of Telecommunications and Ph.D. degrees from the Universidad de Zaragoza in 2005 and 2011 respectively. He is currently a researcher at Universidad de Zaragoza, working in the area of mobile systems. In 2008 he joined Aragón Institute of Engineering Research (I3A) of Universidad de Zaragoza, where he has participated in different projects funded by public administrations and by major industrial and mobile companies. Research interests include wireless communications systems.

# Photos

**José Ramón Gállego**

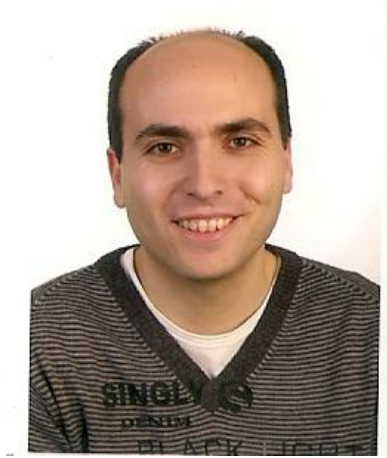

**María Canales**

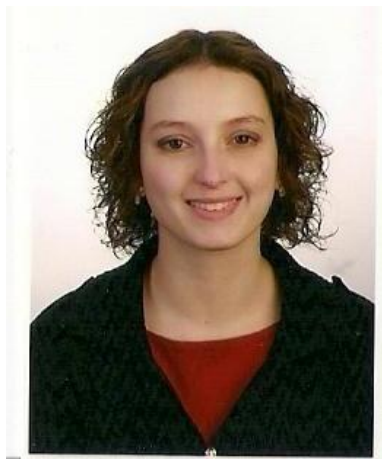

**Jorge Ortín**

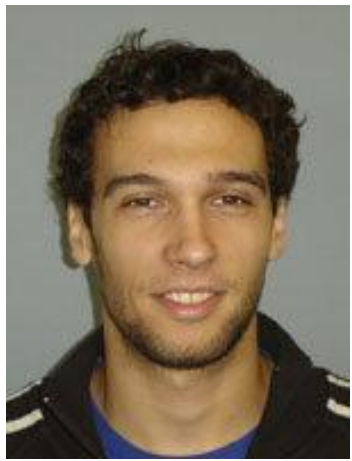